# Accelerating the spin-up of Ensemble Kalman Filtering

Eugenia Kalnay[*] and Shu-Chih Yang
University of Maryland

**Abstract**

A scheme is proposed to improve the performance of the ensemble-based Kalman Filters during the initial spin-up period. By applying the no-cost ensemble Kalman Smoother, this scheme allows the model solutions for the ensemble to be "running in place" with the true dynamics, provided by a few observations.

Results of this scheme are investigated with the Local Ensemble Transform Kalman Filter (LETKF) implemented in a Quasi-geostrophic model, whose original framework requires a very long spin-up time when initialized from a cold start. Results show that it is possible to spin up the LETKF and have a fast convergence to the optimal level of error. The extra computation is only required during the initial spin-up since this scheme resumes to the original LETKF after the "running in place" is achieved.

## 1. Introduction

The relative advantages and disadvantages of 4-dimensional Variational Data Assimilation (4D-Var), already operational in several numerical forecasting centers, and Ensemble Kalman Filter (EnKF), a newer approach that does not require the adjoint of the model, are the focus of considerable current research (e.g., Lorenc, 2003, Kalnay et al, 2007a, Gustafson, 2007, Kalnay et al., 2007b, Miyoshi and Yamane, 2007).

One area where 4D-Var seems to have a clear advantage over EnKF is the initial spin-up, since the evidence thus far is that 4D-Var converges faster than EnKF to its asymptotic level of accuracy. For example, Caya et al. (2005) compared 4D-Var and EnKF for a storm simulating the development in a sounding corresponding to 00UTC 25 May 1999. They found that "Overall, both assimilation schemes perform well and are able to recover the supercell with comparable accuracy, given radial-velocity and reflectivity observations where rain was present. 4DVAR produces generally better analyses than the EnKF given observations limited to a period of 10 min (or three volume scans), particularly for the wind components. In contrast, the EnKF typically produces better analyses than 4DVAR after several assimilation cycles, especially for model variables not functionally related to the observations." In other words, for the severe storm problem the EnKF eventually yields better results than 4D-Var, presumably because of the assumptions made on the 4D-Var background error covariance, but during the crucial *initial time of storm development*, when radar data starts to become available, EnKF provides a worse analysis. For a global shallow water model, which is only mildly chaotic, Zupanski et al. (2006) found that initial perturbations that had horizontally correlated errors converged faster and to a lower level of error than perturbation made with white noise. In agreement with these results, Liu (2007) found using the SPEEDY global primitive equations model that perturbations obtained from differences between randomly chosen states (which are naturally balanced and have horizontal correlations of the order of the Rossby radius of deformation) converged faster than white noise perturbations.

---

[*] Corresponding author: 3431 Computer and Space Sciences Bldg., College Park, MD, 20742-2425. ekalnay@atmos.umd.edu

Yang et al (2008a) compared 4D-Var and the Local Ensemble Transform Kalman Filter (LETKF, Hunt et al., 2007) within a quasi-geostrophic channel model. They found that if the LETKF is initialized from randomly chosen fields, it takes more than 100 days before it converges to the optimal level of error. If, on the other hand, the ensemble mean is initialized from an existent 3D-Var analysis, which is already close to the true state, the LETKF converges to its optimal level very quickly, within about 5 days. However, 4D-Var converges even faster without the need of a good initial guess. This faster spin-up has been observed before and has been attributed to the "smoothing" characteristics of 4D-Var, where the solution fits all the observations within an assimilation window (e.g., Caya et al., 2005, Jidong Gao, 2008, personal communication).

The option of initializing the EnKF from a state close enough to the optimal analysis, such an existent 3D-Var analysis, with balanced perturbations having realistic horizontal correlations, is feasible within a global operational system so that spin-up is not a serious problem. However, there are other situations, such as the storm development discussed above, where radar information is not available before the storm starts, so that no information is available to guide the EnKF in the spin-up towards the optimal analysis. The system may start from an unperturbed state without precipitation, and if a severe storm develops within a few minutes and the EnKF takes considerable real time to spin-up from the observations, it will "miss the train" and give results that are less useful for severe storm forecasting than 4D-Var.

In this note we propose a new method to accelerate the spin-up of the EnKF by "running in place" during the spin-up phase. We find that it is possible to accelerate the convergence of the EnKF so that (in terms of real time) it spins-up even faster than 4D-Var. Section 2 contains a brief theoretical motivation and an introduction of the method, results are presented in Section 3 and a discussion is given in Section 4.

**2. Spin-up, no-cost smoothing and "running in place" in EnKF**

Hunt et al. (2007) provided a derivation of the linear Kalman Filter equations by showing that in the cost function

$$J(\mathbf{x}) = \left[\mathbf{x} - \bar{\mathbf{x}}_n^b\right]^T \left(\mathbf{P}_n^b\right)^{-1} \left[\mathbf{x} - \bar{\mathbf{x}}_n^b\right] + \left[\mathbf{y}_n^o - \mathbf{H}_n\mathbf{x}\right]^T \left(\mathbf{R}_n^{-1}\right)\left[\mathbf{y}_n^o - \mathbf{H}_n\mathbf{x}\right], \qquad (1)$$

the background term represents the Gaussian distribution of a state with the maximum likelihood trajectory (history), i.e., the analysis/forecast trajectory that best fits the data from $t = t_1, ..., t_{n-1}$. This state is obtained by using the forecast model $\mathbf{M}_{t_{n-1}, t_n}$ to advance the previous maximum likelihood analysis $\bar{\mathbf{x}}^a_{n-1}$ and the corresponding analysis error covariance $\mathbf{P}^a_{n-1}$ to the new analysis time $t_n$. In other words, the following relationship is satisfied for some constant $c$:

$$\sum_{j=1}^{n-1}\left[\mathbf{y}^o_j - \mathbf{H}_j \mathbf{M}_{t_n,t_j}\mathbf{x}\right]^T \mathbf{R}^{-1}_j \left[\mathbf{y}^o_j - \mathbf{H}_j \mathbf{M}_{t_n,t_j}\mathbf{x}\right] = \left[\mathbf{x}-\bar{\mathbf{x}}^b_n\right]^T \left(\mathbf{P}^b_n\right)^{-1}\left[\mathbf{x}-\bar{\mathbf{x}}^b_n\right] + c \qquad (3)$$

After the cost function in (1) is minimized finding the analysis $\bar{\mathbf{x}}^a_n$ and its corresponding covariance $\mathbf{P}^a_n$, a similar relationship holds for the analysis at $t_n$ for some constant $c'$:

$$\left[\mathbf{x}-\bar{\mathbf{x}}^b_n\right]^T\left(\mathbf{P}^b_n\right)^{-1}\left[\mathbf{x}-\bar{\mathbf{x}}^b_n\right] + \left[\mathbf{y}^o_n - \mathbf{H}_n\mathbf{x}\right]^T\left(\mathbf{R}^{-1}_n\right)\left[\mathbf{y}^o_n - \mathbf{H}_n\mathbf{x}\right] = \left[\mathbf{x}-\bar{\mathbf{x}}^a_n\right]^T\left(\mathbf{P}^a_n\right)^{-1}\left[\mathbf{x}-\bar{\mathbf{x}}^a_n\right] + c' \qquad (4)$$

Equating the terms in (4) that are linear and quadratic in $\mathbf{x}$, the linear Kalman Filter equations for a perfect model are obtained.

This derivation makes clear that Kalman Filter yields the maximum likelihood estimate $\bar{\mathbf{x}}^a_n$ with the corresponding error covariance $\mathbf{P}^a_n$ at time $t_n$ if the model is perfect and *if the previous analysis $\bar{\mathbf{x}}^a_{n-1}$ at $t_{n-1}$ is also the maximum likelihood estimate at the previous analysis time*. Hunt et al. (2007) also indicate that a system can be initialized with a limited number of observations at the initial time $t_1$ by assuming that the initial background error covariance is large but not infinitely large. Although the initial cost function has an additional quadratic term, they point out that "with sufficient observations over time, the effect of this term [on the background error covariance] at time $t_n$ decreases in significance as n increases". In other words, with sufficient observations, the Kalman Filter eventually converges and yields the maximum likelihood solution and its error covariance.

The EnKF also provides a maximum likelihood analysis, except that the background and analysis error covariances are estimated from an ensemble of K generally nonlinear forecasts:

$$\mathbf{P}^b_n \approx \frac{1}{K-1}\mathbf{X}^{bT}_n \mathbf{X}^b_n, \qquad (5)$$

where $\mathbf{X}^b_n$ is a matrix whose columns or the background (forecast) perturbations $\mathbf{x}^b_{n,k} - \bar{\mathbf{x}}^b_n$ and $\bar{\mathbf{x}}^b_n$ is the most likely forecast state, i.e., the ensemble average. Similar equations are valid for the analysis mean $\bar{\mathbf{x}}^a_n$ and the analysis error covariance $\mathbf{P}^a_n$.

EnKF, like Kalman Filter, is thus a sequential data assimilation system where, after the new data is used at the analysis time, it should be discarded, but this is true only if the previous analysis and the new background are the most likely states given the past observations. In other words, *if the system has converged after the initial spin-up all the information from past observations is already included in the background*. In contrast, 4D-Var is a smoother that best fits all the observations (even asynoptic data) within an assimilation window. We note that EnKF can be also easily extended to 4-dimensions as in 4D-Var, allowing for the assimilation of asynoptic observations made between two analyses (Hunt et al., 2004). In EnKF only the observational increments that project on the subspace of the ensemble forecasts can be assimilated. Therefore the observational increments computed at the observation time, which are linear combinations of the ensemble forecasts, can be moved forward (or backward) to the analysis time by simply using the same linear combination of the ensemble forecasts obtained at the observation time.

In summary, after the initial spin-up all the information from past observations is already included in the background field, so that the observations should be used only once and then discarded. However, there is no theoretical reason why this restriction should also be applied when EnKF is "cold-started". In practical applications, nevertheless, the rule of using the data only once is usually still applied (e.g., Zupanski et al. 2006), and a slow EnKF spin-up is usually observed. In this note we suggest that when a quick EnKF spin-up (in real time) is needed in order to make useful short-range forecasts for fast weather instabilities, the initial observations can be used more than once in order to extract more information from them, and that this procedure leads to a much faster spin-up of the initial ensemble in real time. This "running in place" algorithm is made possible by the use of a "no-cost" Ensemble Kalman Smoother (EnKS) proposed by Kalnay et al. (2007b) and tested by Yang et al. (2008a).

The no-cost EnKS is simple and easy to implement. Consider an assimilation window $[t_{n-1}, t_n]$ within a Square-Root type of EnKF (e.g., Tippett et al. 2003, Whitaker and Hamill, 2002, Ott et al., 2004). The analysis ensemble members at time $t_n$ are each a weighted average (linear combination) of the ensemble forecasts at $t_n$ (Hunt et al., 2007). Yang et al. (2008b) explored the characteristics of these analysis weight fields and found that they vary smoothly on large scales. As a result, if the analysis (i.e., the computation of the weights) is carried out on a very sparse analysis grid and then interpolated to the in-between grid points, the interpolated weight analysis is not only computationally more efficient, but the interpolation does not degrade and may actually improve upon the full resolution analysis.

The no-cost EnKS is obtained by simply applying the same weights obtained at analysis time $t_n$ to the initial ensemble at $t_{n-1}$ (Kalnay et al., 2007b). Yang et al. (2008a) tested this scheme and found that indeed, the no-cost EnKS smoothed ensemble at $t_{n-1}$ is more accurate than the analysis ensemble valid at $t_{n-1}$, as could be expected from the fact that the smoothed ensemble at the beginning of the window has benefited from the information provided by the "future" observations in the window $[t_{n-1}, t_n]$. Although the no-cost smoothing improves the initial analysis at $t_{n-1}$, it does not improve the final analysis at $t_n$, since the forecasts started from the new initial analysis ensemble will end as the final analysis ensemble (at least in a linear sense).

With the no-cost EnKS it is then possible to use the initial observations repeatedly in order to extract maximum information from the observations and improve the quality (likelihood) of the initial ensemble faster, which leads the ensemble-based background error covariance to be more representative of the true forecast error statistics.

The algorithm that we have tested is as follows: We start the EnKF from poor initial ensemble mean and random perturbations at $t_0$, and integrate the initial ensemble to $t_1$. Then the "running in place" loop with $n=1$, is:

a) Perform a standard EnKF analysis and obtain the analysis weights at $t_n$, saving the mean square observations minus forecast (OMF) that is computed by the EnKF.

b) Apply the no-cost smoother to obtain the smoothed analysis ensemble at $t_{n-1}$ by using the same weights obtained at $t_n$.

c) Perturb the smoothed analysis ensemble with a small amount of random Gaussian perturbations, a method similar to additive inflation. These added perturbations have two purposes: they avoid the problem of otherwise reaching the same final analysis at $t_n$ as in the previous iteration, and they allow the ensemble perturbations to evolve into fast growing directions that may not have been included in the unperturbed ensemble subspace.

d) Integrate the perturbed smoothed ensemble to $t_n$. If the forecast fit to the observations is smaller than in the previous iteration according to a criterion such as

$$\frac{OMF^2(iter) - OMF^2(iter+1)}{OMF^2(iter)} > \varepsilon, \qquad (6)$$

go to a) and perform another iteration. If not, let $t_{n-1} \leftarrow t_n$ and proceed to the next assimilation window.

### 3. Results

Figure 1 shows the spin-up obtained using several methods over 200 analysis cycles of 12 hours each (corresponding to 100 days), all starting from a randomly chosen mean state and from perturbations created as Gaussian noise. The standard LETKF (black line) is then run forward as is conventionally done, using the observations only once. It takes over 120 cycles for the ensemble perturbations to "breed" into the "errors of the day", and between 120 and 170 cycles the LETKF converges rather quickly to the optimal level of error. The 4D-Var (blue line) with the same initial guess and 12-hour windows starts spinning down immediately and after 80 cycles it has already converged to its optimal level. After they attain convergence, LETKF and 4D-Var errors compared to the true state are similar.

We performed a preliminary experiment allowing for repeated use of the observations by fixing the number of "running in place" iterations at 10 (dashed black line). The LETKF with 10 iterations spins-down even faster than 4D-Var and converges in about 50 cycles but to a level of error much higher than optimal. This is not surprising, since once the system is close to the maximum likelihood solution, as indicated by the theoretical arguments discussed above, observations should be used only once and then discarded.

The adaptive approach (6) tests whether the system is optimal by checking whether iterations reduce the ensemble forecast error, and stops iterating when the relative improvement is less than $\varepsilon$. Figure 1 suggests that a low value of $\varepsilon = 0.01$ (associated with a larger number of iterations, as shown in Figure 2, leads to a faster (but costlier) initial spin-down of the error, but that after convergence it is less than optimal because it requires too many iterations (thin red line with crosses). Values of $\varepsilon$ between 0.02 (not shown) and 0.05 (thick red line) are optimal since they lead to a spin-down of the initial errors faster than 4D-Var, and a final level of error at least as good as that of 4D-Var.

Finally, we tested whether the use of additive perturbations that have horizontal correlations accelerates the spin-up, as found by Zupanski et al. (2006) for the initial perturbations. Figure 1 shows the result of the LETKF when the additive perturbations are chosen so that their background error covariance is the 3D-Var covariance (green line), i.e., the columns of the matrix $\sqrt{\mathbf{B}_{3D-Var}}\mathbf{E}$, where $\mathbf{E}$ is a matrix whose columns are random Gaussian numbers such that $\mathbf{EE}^T = \mathbf{I}$. Since $\mathbf{B}_{3D-Var}$ was obtained using the NMC method (Parrish and Derber, 1992, Yang et al., 2008a), the perturbations based on $\mathbf{B}_{3D-Var}$ have horizontal correlation lengths with synoptic scales, whereas the additive noisy perturbations used for the other experiments discussed before have very small correlation lengths.

Figure 1 shows that when the additive perturbations are horizontally correlated (as in the green line corresponding to $\mathbf{B}_{3D-Var}$ perturbations), convergence takes place faster than with the noisy additive perturbations, even when the same criterion $\varepsilon = 0.05$ is used for both. This agrees well with the conclusion of Zupanski et al. (2006) that horizontal correlation of the perturbations accelerates spin-up. Nevertheless, once convergence has been achieved, the accuracy of the system with noisy perturbations (red) is slightly better than the system with $\mathbf{B}_{3D-Var}$ perturbations.

Figure 2 compares the number of iterations required by all the "running in place" schemes presented in Figure 1. It shows that the number of iterations required with $\varepsilon = 0.01$ starts with about 50, and remains at a range of 2-10 iterations even after convergence, suggesting that the criterion is too strict, leading to inefficient spin-up. With $\varepsilon = 0.05$ the system with synoptic scale ($\mathbf{B}_{3D-Var}$-based) additive perturbations converges faster, reaching 1-2 iterations after only about 30 data assimilation cycles, and then oscillates between 1 and 2 iterations. The system with noisy, small scale additive inflation (also with $\varepsilon = 0.05$) takes about 50 data assimilation cycles to reach 1 iteration, but then remains at that level.

## 4 Discussion

The results obtained are very encouraging: it is possible to spin-up the LETKF (and other EnKF algorithms) when a cold-start and fast convergence to the optimal level of error (in terms of real or physical time) are required, by simply using the initial observations many times rather than once. This is made possible by the no-cost Ensemble Kalman Smoother proposed by Kalnay et al (2007b), where the smoothed analysis ensemble at the beginning of an assimilation window is given by the analysis weights of the ensemble forecast at the end of the window. It is necessary to add perturbations to the ensemble, in a procedure akin to additive inflation. The number of iterations needed is estimated by checking whether the smoothed analysis reduces the forecast error (OMF). A level of relative reduction $\varepsilon$ of about 2-5% was found to work well in this quasi-geostrophic model, leading to about 5-10 initial iterations that are reduced to one or two when the system converges.

Although this spin-up reduction method requires more computations during the initial spin-up, this is only a temporary overhead, and once the number of iterations becomes 2 or less, the "running in place" can be stopped. In the case of a developing storm, it would be possible to use the weight interpolation algorithm of Yang et al (2008b) to perform the additional iterations locally, "where the action is", rather than throughout the whole domain. We explored some parameters such as the reduction of Observation-Minus-Forecast statistics for determining whether more iterations should be performed and found that a criterion of at least 2-5% seems to work well. Similarly, we found that additive inflation with horizontal correlations accelerates the initial spin-up, in agreement with Zupanski et al. (2006), but later is slightly worse. These explorations are only indicative for the quasi-geostrophic model we have used, and other factors may be also important in more realistic situations.

Acknowledgements:

We are grateful to Dr. Jidong Gao who pointed out the difficult problem of slow spin-up of EnKF in severe storm prediction using radar data. The influence of B Hunt, I Szunyogh and E Kostelich paper is also gratefully acknowledged.

## 5. References:

Caya, C, J. Sun, and C. Snyder, 2005: A Comparison between the 4DVAR and the Ensemble Kalman Filter Techniques for Radar Data Assimilation, *Mon. Wea. Rev.*, **133**, 3081-3094.

Hunt, B. R., E. Kostelich, I. Szunyogh, 2007: Efficient data assimilation for spatiotemporal chaos: a Local Ensemble Transform Kalman Filter. *Physica D,* **230***,* 112-126.

Kalnay, E., H. Li, T. Miyoshi, S.-C. Yang and J. Ballabrera, 2007a: 4D-Var or Ensemble Kalman Filter? *Tellus A*, **59**, 758–773.

----------------------------------------------------------------------, 2007b: Response to Comments by N. Gustaffson. *Tellus A*, **59**, 778-780

Liu, J., 2008: Applications of the LETKF to adaptive observations, analysis sensitivity, observation impact and the assimilation of moisture. *PhD dissertation*, University of Maryland, 154 pages.

Lorenc, A. C., 2003: The potential of the ensemble Kalman filter for NWP – a comparison with 4D-Var. Quart. J. Roy. Meteor. Soc., 129, 3183-3203.

Ott, E., B. R. Hunt, I. Szunyogh, A. V. Zimin, E. J. Kostelich, M. Corazza, E. Kalnay, D. J. Patil, and J. A. Yorke, 2004: A local ensemble Kalman filter for atmospheric data assimilation. *Tellus*, **56A**, 415-428.

Parrish, D. and J. Derber, 1992: The National Meteorology Center's spectral statistical-interpolation analysis system. *Mon. Wea. Rev*., **120**, 1747-1763.

Tippett, M. K., J. L. Anderson, C. H. Bishop, T. M. Hamill, and J. S. Whitaker, 2003: Ensemble Square Root Filters. *Mon. Wea. Rev.*, **131**, 1485-1490.

Whitaker, J. S. and T. M. Hamill, 2002: Ensemble data assimilation without perturbed observations, Mon. Wea. Rev. 130, 1913–1924.

Yang, S-C, M. Corazza, A. Carrassi, E. Kalnay, and T. Miyoshi, 2008a: Comparison of ensemble-based and variational-based data assimilation schemes in a quasi-geostrophic model. *Mov. Wea. Rev., under revision*.

Yang, S-C, E. Kalnay, B. Hunt, N. Bowler, 2008b: Weight interpolation for efficient data assimilation with the Local Ensemble Transform Kalman Filter, *Quart. J. Roy. Meteor. Soc.*, *under revision*.

Zupanski, M., S. J. Fletcher, I. M. Navon, B. Uzunoglu, R. P. Heikes, D. A. Randall, T. D. Ringlee and D. Daescu, 2006: Initiation of ensemble data assimilation. Tellus, 58A, 159-170.

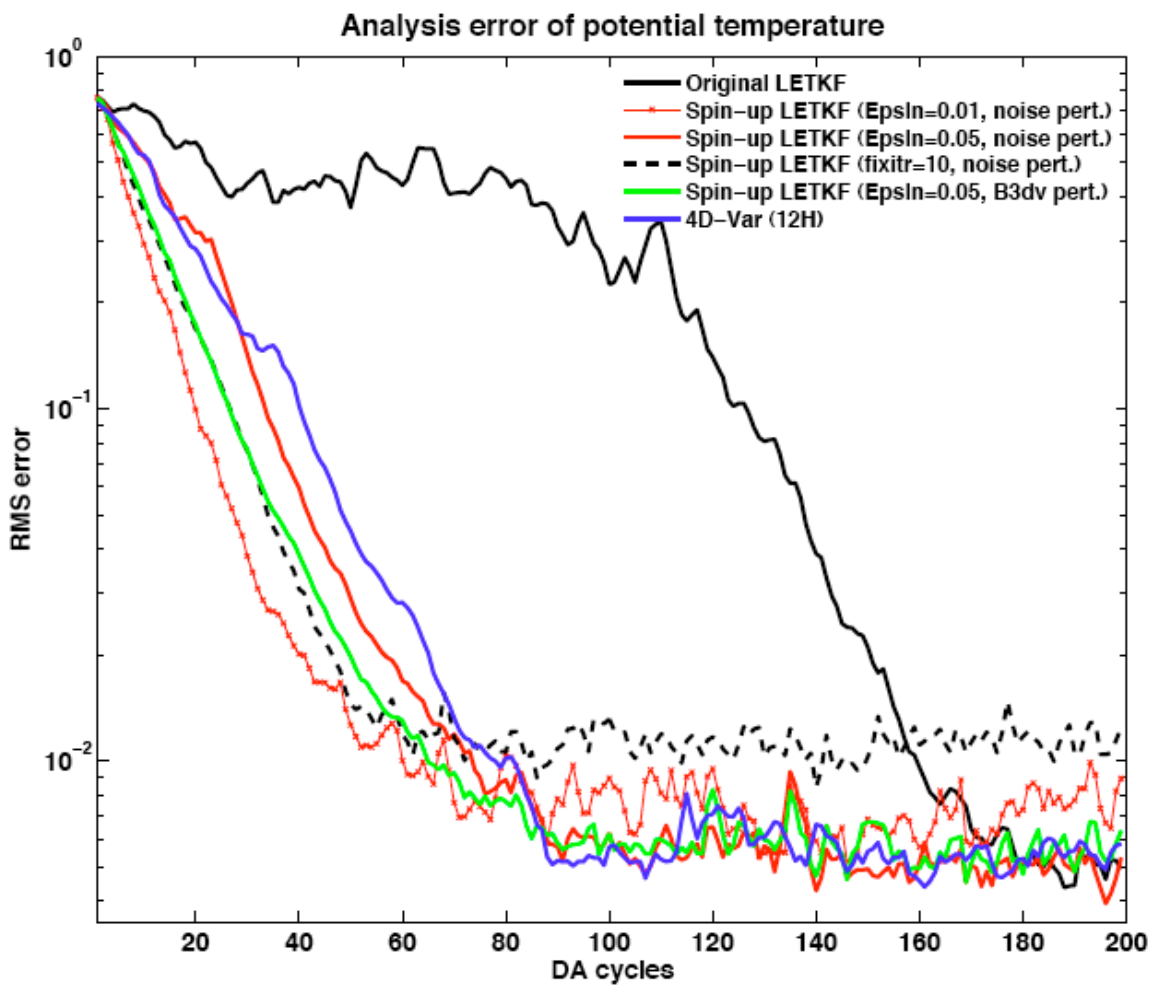

**Figure 1** Time series of RMS analysis errors in potential temperature at the bottom level of the original LETKF (black line), the spin-up LETKF with Gaussian noises and epsilon=0.01 (thin red line with ×), with Gaussian noises and epsilon=0.05 (red line), with Gaussian noises and 10 iterations (black dashed line), with 3D-Var noises and epsilon=0.05 (green line) and 4D-Var (blue line).

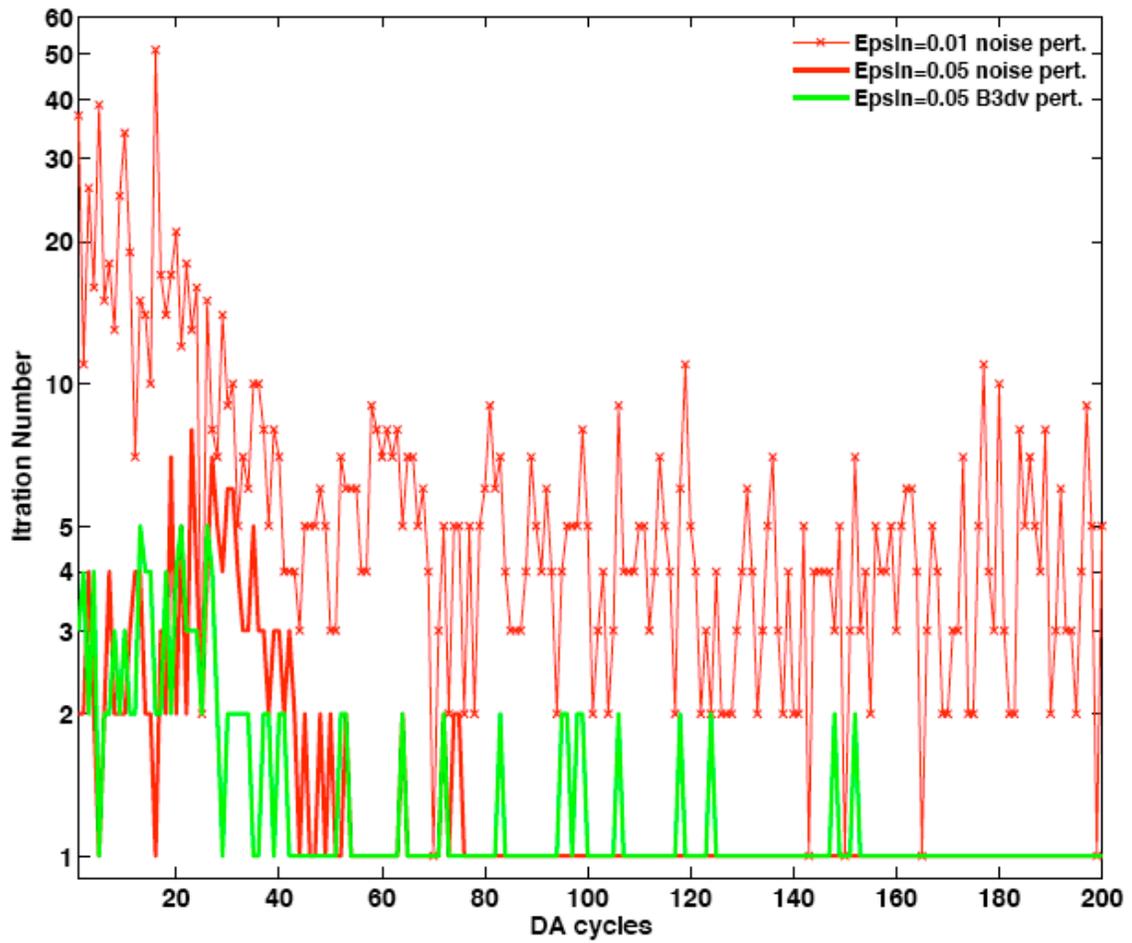

**Figure 2** The number of iteration required by the spin-up LETKF with Gaussian noises and epsilon=0.01 (thin red line with ×), with Gaussian noises and epsilon=0.05 (red line) and with 3D-Var noises and epsilon=0.05 (green line)